\documentclass[aps,prl,twocolumn,superscriptaddress,amsmath,amssymb]{revtex4-2}

\usepackage[utf8]{inputenc}
\usepackage{amsmath,amsthm}
\usepackage{graphicx}
\usepackage[colorlinks = true,
            linkcolor = blue,
            urlcolor  = blue  ,
            citecolor = blue,
            anchorcolor = blue]{hyperref}
\usepackage{braket}
\usepackage{xcolor}
\newcommand{\real}{{\rm Re}}

\newcommand{\bea}{\begin{eqnarray}}
\newcommand{\eea}{\end{eqnarray}}

\newcommand{\bfvec}[1]{\mathbf{#1}}

\global\long\def\ga{\gamma} \global\long\def\de{\delta}

\global\long\def\ell#1{\theta_{#1}}

 \global\long\def\ka{\kappa}
\global\long\def\si{\sigma}
\global\long\def\vfi{\varphi}
\global\long\def\s{\sigma}

 \global\long\def\Om{\Omega}
\global\long\def\eps{\epsilon}
\global\long\def\al{\alpha}

\global\long\def\ga{\gamma} \global\long\def\de{\delta}

\global\long\def\no{\nonumber}

\theoremstyle{thm@}

\theoremstyle{remark}

\global\long\def\braket#1#2{\left\langle #1|#2\right\rangle }

\def\pp{\mathbf{p}}
\def\qq{\mathbf{q}}

\newcommand{\kind}{l}

\let\Ph=\phi
\let\PH=\Phi
\def\i{{\rm i}}
\DeclareMathOperator{\re}{e}
\let\e=\varepsilon
\theoremstyle{plain}

\newtheorem*{theorem*}{Theorem}
\def\pv{\mathbf{p}}
\def\qv{\mathbf{q}}
\def\Fv{\mathbf{F}}
\def\Hv{\mathbf{H}}
\def\epc{,}
\def\epp{.}
\DeclareMathOperator{\sign}{sign}
\let\p=\pi
\def\rd{{\rm d}}

\usepackage{amsfonts}
\usepackage{amsmath}
\usepackage{amsthm}
\usepackage{amssymb}
\usepackage{stmaryrd}
\usepackage{amscd}
\usepackage{eucal}
\usepackage{dsfont}
\usepackage{bm}
\usepackage{graphicx}

\allowdisplaybreaks

\begin{document}

\title{Chiral basis for qubits and spin-helix decay}
\author{Vladislav Popkov}
 \affiliation{Department of Physics,
  University of Wuppertal, Gaussstra\ss e 20, 42119 Wuppertal,
  Germany}
\affiliation{Faculty of Mathematics and Physics, University of Ljubljana, Jadranska 19, SI-1000 Ljubljana, Slovenia}
\author{Xin  Zhang}
\affiliation{Beijing National Laboratory for Condensed Matter Physics, Institute of Physics, Chinese Academy of Sciences, Beijing 100190, China}
\author{  Frank G\"ohmann} 
\affiliation{Department of Physics,
 University of Wuppertal, Gaussstra\ss e 20, 42119 Wuppertal,
  Germany}
\author{  Andreas Kl\"umper} 
\affiliation{Department of Physics,
 University of Wuppertal, Gaussstra\ss e 20, 42119 Wuppertal,
  Germany}

\begin{abstract}
We propose a qubit basis composed of transverse spin helices with
kinks. Unlike the usual computational basis, this chiral basis
is well suited for describing quantum states with nontrivial
topology. Choosing appropriate parameters the operators of the
transverse spin components, $\sigma_n^x$ and $\sigma_n^y$, become
diagonal in the chiral basis, which facilitates the study of problems
focused on transverse spin components. As an application, we study
the temporal decay of the transverse polarization of a spin helix
in the XX model that has been measured in recent cold atom experiments.
We obtain an explicit universal function describing the relaxation
of helices of arbitrary wavelength.
\end{abstract}
\maketitle

\textbf{Introduction--} 
A proper choice of basis often is the crucial first step toward
success. For example, the modes of the harmonic oscillator are
best described by the coherent state basis. The use of wavelets
is well suited for describing signals confined in space or time
\cite{Wavelets}, while the Fourier basis is natural for solving
linear differential equations with translational invariance.

For qubits, i.e., quantum systems with spin-$1/2$ local degrees
of freedom, the most widely used basis is the computational basis,
which is composed of tensor products of the eigenstates $\binom{1}{0}$,
$\binom{0}{1}$ of the $\sigma^z$ operator. The advantages of the
computational basis are its factorized structure, orthonormality,
and $U(1)$-symmetry `friendliness'. The computational basis is
well-suited for calculating spectra and correlation functions
of local operators for Hamiltonians (like that of the XXZ model)
that preserve the total magnetization in $z$-direction.

However, the computational basis appears poorly equipped to
describe states with nontrivial topology, such as chiral states, current-carrying
states, or states with windings. One prominent example is the spin-helix
state in a 1-dimensional spin chain,
\begin{equation} \label{def:SHS}
     \ket{\Psi(\al_0,\eta)} = \bigotimes_n \ket{\phi(\al_0 + n\eta)},
\end{equation}
where $\ket{\phi(\al)}$ describes the state of a qubit, while $+n\eta$
represents the linear increase of the qubit phase along the chain,
parameterized in a proper model-dependent way. Thanks to their
factorizability, spin helices (\ref{def:SHS}) are straightforward
to prepare in experimental setups that allow for an adjustable spin
exchange, such as those involving cold atoms
\cite{SHS-Ketterle,2020NatureSpinHelix,2021KetterleTransverse}.
These helices possess interesting properties as evidenced by both
experimental
\cite{SHS-Ketterle,2020NatureSpinHelix,2021KetterleTransverse}
and theoretical \cite{SHS-Phantom,Phantom-Long,ChiralBA,SHS-Hydro}
studies. It was suggested that quantum states with helicity are
even better protected from noise than the ground state, and that the
helical protection extends over intermediate timescales \cite{2023Posske}.

Since the spin helix state (\ref{def:SHS}) is not an eigenstate
of the operator of the total magnetization, it is not confined to
a single $U(1)$ block, but is given by a sum over all the blocks,
with fine-tuned coefficients, as shown in (\ref{eq:SHSinU(1)}) and
(\ref{eq:PBRstates}), even for spatially homogeneous spin helices
($\eta=0$ in (\ref{def:SHS})). A simple shift of the helix phase,
$\al_0 \rightarrow \al_0 + const$ in (\ref{def:SHS}), gives a linearly
independent state with the same qualitative properties (winding, current,
etc.). However, to represent such a shift in the standard computational
basis, all the fine-tuned expansion coefficients must be changed
in a different manner.

We shall introduce an alternative basis, all components of which
are chiral themselves and thus ideally tailored for the description
of chiral states. This chiral basis consists of helices and helices
with kinks (phase dislocations). It provides a block hierarchy
based on the number of kinks (rather than on the number of down
spins as in case of the computational basis). Unlike the computational
basis, the chiral basis is intrinsically topological.

We diagonalize the XX Hamiltonian in the chiral basis and apply the
chiral eigenbasis to the problem of spin-helix decay under
XX dynamics. This is a problem of its own relevance for experiment
\cite{SHS-Ketterle,2021KetterleTransverse} and theory. Except for
some quench problems, only very few explicit examples
\cite{1970McCoy,2019Sasamoto,2022Essler} of exact non-equilibrium
dynamics of many-body quantum systems are known. Those examples
rely on the summation of series of matrix elements and overlaps
of initial states with Bethe eigenstates. Although the latter
are known in our spin helix case \cite{JiPo20}, their summation
has remained a problem. The chiral basis helps to circumvent
this problem, as we have other selection rules for the matrix
elements and obtain particularly convenient forms of the overlaps.
Unlike previous works \cite{2022XZ-SHS,SHS-Hydro} which dealt with
the simpler case of spin helices modulated in the $XZ$-plane,
we deal with transverse spin helices modulated in the $XY$-plane.
Longitudinal and transverse spin helices behave rather differently.
In particular, transverse helices can exhibit quantum scars
\cite{2022-QuantumScars-Review} under XXZ dynamics.

\textbf{Chiral multi-qubit basis--}
Our starting point is the `winding number operator'
\begin{equation} \label{eq:Uoper}
     V = {\textstyle \frac{1}{4}}
         \sum_{k=1}^{N/2} \left(\si_{2k-1}^x \si_{2k}^y - \si_{2k}^y \si_{2k+1}^x\right),
\end{equation}
defined for an even number of qubits $N$. It has remarkably
simple factorized eigenstates. A state 
\begin{equation} \label{eq:Psi}
     \Psi = 2^{-N/2} \vfi_1 \otimes \zeta_1 \otimes  \vfi_2 \otimes \zeta_2
                            \otimes \ldots  \otimes \vfi_{N/2} \otimes \zeta_{ N/2}  
\end{equation}
is an eigenstate of $V$, if all odd (even) qubits are polarized
in  positive or negative $x$- ($y$-) direction,
\begin{equation} \label{eq:Psi1}
     \bra{\vfi_j} = (1,\pm 1), \quad \bra{\zeta_j} = (1,\mp \i). 
\end{equation}
In such states the qubit polarization at each link between
$n$, $n+1$ changes by an angle of $+\pi/2$ or $-\pi/2$ in
the XY-plane. Each anticlockwise or clockwise rotation by
$\pi/2$ adds $+1$ or $-1$ to the eigenvalue of $4V$ so that  
\begin{equation}
     V \Psi = {\textstyle \frac{1}{4}} (N-2M) \Psi,
\end{equation}
where $M$ is the number of clockwise rotations, further
referred to as \textit{kinks}. Clearly, every state $\Psi$
in (\ref{eq:Psi}) is uniquely characterized by the kink
positions $1 \le n_1 < \dots < n_M \le N$ (between qubits $n_k$,
$n_k+1$), and the polarization $\kappa = \pm$ of the first
qubit $\vfi_1$. We denote this state by $\i^{\sum_{k} n_k}
\ket{\ka; {\bf n}}$, where ${\bf n} = (n_1, \dots, n_M)$.
Then, by construction, the set of $V$ eigenstates 
\begin{equation} \label{eq:PsiBasis}
     \bigl\{\ket{\ka; \bf{n}}\big| \kappa = \pm,
            1 \le n_1 < \dots < n_M \le N\bigr\}
\end{equation}
is an orthonormal basis of $N$ qubits that we call the chiral
basis. For compatibility with periodic boundary conditions
the winding number $(N - 2M)/4$ must be an integer, implying
that $M$ must be even (odd) if $N/2$ is even (odd). We shall 
call such values of $M$ admissible.

\textit{Remark 1.} The chiral basis vectors have a topological
nature; a single kink cannot be added to (removed from)
a periodic chain by the action of a local operator. In an open
chain this is only possible at the boundary.

\textit{Remark 2.} Applying $\sigma_n^z$ in a kink-free zone creates a kink
pair at the neighbouring positions $n-1$ and $n$.  Applying a
string of operators $\sigma_n^z \sigma_{n+1}^z \ldots \sigma_{n+k}^z$ in a
kink-free zone creates two kinks at a distance of $k+1$,  e.g.,
\begin{equation}
     \ket{+;1,k+2} = \si_2^z \ \si_3^z \ldots  \si_{k+2}^z \ket{+},
\end{equation}
where $\ket{+}$ is a perfect spin helix of type (\ref{def:SHS}),
\begin{equation} \label{def:vac}
     \ket{+} = \ket{\rightarrow \uparrow \leftarrow
                    \downarrow \rightarrow \uparrow \leftarrow \downarrow \ldots},
\end{equation}
i.e., a spin helix with maximal winding number $N/4$. Here the
arrows depict the polarization of the qubits in the XY-plane,
e.g., $\uparrow,\downarrow $ depict a qubit $\zeta_j$ (\ref{eq:Psi1})
with polarization along the $y$ axis.

\textit{Remark 3.} The connection between the chiral basis and the standard
computational basis is nontrivial. For example, the chiral vacuum state (\ref{def:vac})
is expanded in terms of the computational basis as 
\begin{align}
&\ket{+}=2^{-\frac{N}{2}}\sum_{n=0}^N (-\i)^n\xi_n,\label{eq:SHSinU(1)}\\
&\xi_n\!=\!\frac{1}{n!}\, \sum_{\kind_1,\ldots ,\kind_n=1}^{N}\,
\i^{\kind_1+\ldots +\kind_n}\,
\si_{\kind_1}^{-}\ldots  \si_{\kind_n}^{-}\, \binom{1}{0}^{\otimes_N},\label{eq:PBRstates}
\end{align}
see \cite{SHS-Phantom} for a proof.

In the following we will explore two applications of the chiral basis
that are related. First, we will use it to classify the eigenstates
of the XX model according to the number of kinks.

\textbf{Eigenstates of the XX model within the chiral sectors--}
The crucial observation is that  $V$ commutes with the Hamiltonian of the XX model,
\begin{equation}
H=\sum_{n=1}^{N} \si_n^x \si_{n+1}^x+ \si_n^y \si_{n+1}^y,
\quad \vec{\sigma}_{N+1}\equiv \vec{\sigma}_1. \label{def:XX0}
\end{equation}
Consequently, $H$ is block diagonal in the chiral basis with
each block corresponding to a fixed number of kinks $M$. For
one-kink states $M=1$ we obtain
\begin{align*}
&H \ket{\ka;n} = 2 \ket{\ka;n-1}+ 2 \ket{\ka;n+1}, \quad n \neq 1,N, \\
&H \ket{\ka;1} = -2 \ket{-\ka;N}+ 2 \ket{\ka;2}, \\
&H \ket{\ka;N} = -2 \ket{-\ka;1}+ 2 \ket{\ka;N-1}. 
\end{align*}
The $2N$ eigenstates of $H$ belonging to the one-kink 
subspace are given by the ansatz
\begin{align}
& \ket{\mu_1(p)}= \frac{1}{\sqrt{2N}}\sum_{n=1}^N  \re^{\i p n} \left(\ket{+;n}-\re^{\i p N} \ket{-;n}\right),  \label{eq:kink1Eigen}\\
&\re^{\i p N} = \pm 1,
\end{align}
where $p$ is a chiral analogue of the quasi-momentum.  
The diagonalization of $H$ within a subspace with an arbitrary
number of kinks can be performed by the coordinate Bethe Ansatz
(see \cite{Supp2}) which gives a complete set of eigenvectors in
the chiral basis.
\begin{theorem*}
The states
\begin{align}
     |\mu_M(\pv)\rangle & = \sum_{1 \le n_1 < \ldots < n_M \le N}
        \chi_{\bf n} (\pv) \bigl\{|1;{\bf n}\rangle - \re^{\i p_1 N} |-1;{\bf n}\rangle\bigr\},
	   \notag \\
        \chi_{\bf n} (\pv) & = \frac{1}{\sqrt{2 N^M}} \,
                        \det_{j,k = 1, \dots, M} \bigl\{\re^{\i p_j n_k}\bigr\}, \notag \\
     |u; {\bf n}\rangle & = (-\i)^{\sum_{j=1}^M n_j} \bigotimes_{k=1}^{n_1}
                  \psi_k(u)\bigotimes_{k=n_1+1}^{n_2} \psi_k(u+2) \notag
		  \\ & \phantom{xxxxxxxxxxxxxxxxx} \cdots
		  \bigotimes_{k = n_M + 1}^{N} \psi_k(u+2M), \notag \\ \label{eq:Mshock}
     \psi_k(u) & = \frac{1}{\sqrt{2}}\binom {1}{\re^{\frac{\i  \pi}{2}(k- u)}},
\end{align}
where $M$ is admissible and where the chiral quasi-momenta $\pv =
(p_1, p_2, \ldots,p_M)$ satisfy either $\re^{\i p_j N} = 1$ or
$\re^{\i p_j N} = -1$ for all $p_j$, form an orthonormal basis of
eigenstates of Hamiltonian (\ref{def:XX0}),
$\langle\mu_M (\pv)|\mu_{M'}(\pv')\rangle = \de_{\pv,\pv'} \de_{M,M'}$.
The corresponding energy eigenvalues are $E_{\pv} = \sum_{j=1}^{M} \e_{p_j}$,
$\e_p = 4 \cos (p)$.
\end{theorem*}

Some clarifications might be appropriate here. First, the states (\ref{eq:Mshock})
are a generalization of those in (\ref{eq:PsiBasis}) by an additional rotation of
all qubits by the same angle $\pi (1 -u)/2$ in the XY-plane. Setting $u=1$ yields
(\ref{eq:PsiBasis}). The extra degree of freedom originates from the $U(1)$
symmetry of the XX model.

Second, the XX eigenstates in the chiral basis formally resemble those in the usual
computational basis \cite{1993ColomoXX0basis}, where the number of spins up plays
the role of the number of kinks. In particular, the wave functions
$\chi_{\bf n} (\pv)$ have the familiar form of Slater determinants.

\textbf{Spin-helix  decay in the XX model--}
Next, we apply our chiral basis to  study the time evolution of a
transverse spin-helix magnetization profile, measured in
\cite{SHS-Ketterle}. We are able to obtain an exact and explicit
answer, when the time evolution of the local spin $\vec \si_n (t) =
\re^{\i H t} \vec \si_n \re^{- \i H t}$ is driven by the XX Hamiltonian
(\ref{def:XX0}). The initial spin helix in the XY-plane is described
by the state
\begin{equation}
     \ket{\Psi_Q} = \frac{1}{\sqrt{2^N}}\bigotimes_{n=1}^{N}
                    \begin{pmatrix} \re^{- \frac{\i nQ}{2}} \\
		                    \re^\frac{\i nQ}{2} \end{pmatrix},
\end{equation}
where the wave vector $Q$ satisfies the commensurability condition
$QN = 0 \mod 2 \pi$. Setting $\PH = \sum_{n=1}^N Qn \si_n^z$ and
$\ket{\Om} = \ket{\Psi_0}$ we see that $\ket{\Psi_Q} =
\re^{- \frac{\i \PH}{2}} \ket{\Om}$. The operator
$\re^{- \frac{\i \PH}{2}}$ induces a rotation of qubits at site
$n$ about the $z$-axis by an angle $Qn$. Thus,
\begin{equation} \label{initialhelix}
     \bra{\Psi_Q} \vec \si_n \ket{\Psi_Q} = 
     \bra{\Om} \re^\frac{\i \PH}{2} \vec \si_n \re^{- \frac{\i \PH}{2}} \ket{\Om} = 
     \begin{pmatrix} \cos(Qn) \\ \sin(Qn) \\ 0 \end{pmatrix}
\end{equation}
which is precisely a spin helix in the XY-plane. In order to
obtain its time evolution we have to replace $\vec \si_n$
by $\vec \si_n (t)$ in (\ref{initialhelix}) and commute
$\re^\frac{\i \PH}{2}$ through $H$. The details of this
calculation can be found in \cite{2023SHSdecay,Supp2}. We
finally obtain a remarkably simple structure:
\begin{align} \label{eq:Sx}
     & \bra{\Psi_Q} \vec \si_n (t) \ket{\Psi_Q}
        = S_N \bigl(\cos(Q) t \bigr)
            \begin{pmatrix} \cos(Qn) \\ \sin(Qn) \\ 0 \end{pmatrix}, \\
     & S_N (t) = \bra{\Om} \si_1^x (t) \ket{\Om}.
     \label{eq:SN(t)}
\end{align}
The shape of the magnetization profile is not changing with time.
The profile fades away with an amplitude $S_N (\cos(Q) t)$ which
depends on the wave vector $Q$ in a self-similar way.

The remaining universal function $S_N (t)$ is still a
non-equilibrium one-point function and therefore hard
to calculate. Free fermion techniques involving a
Jordan-Wigner transformation do not seem to sufficiently
simplify the problem, because the  density matrix associated
with $\ket{\Omega}$ is not an exponential of bilinear
expressions in Fermi operators $c_j,c_j^\dagger$, see
\cite{2022AresCalabrese}. As previously seen in other
examples \cite{1993ColomoXX0basis,GKS20a} an appropriately
adapted Bethe Ansatz technique, in the present case
based on the chiral basis, turns out to be more efficient.

A key simplification in calculating $S_N(t)$ by means of
the chiral basis consists in the fact that, with $u=1$ in
(\ref{eq:Mshock}), the operator $\si_1^x$ becomes diagonal
in the chiral basis
\begin{equation} \label{eq:SigmaxAction}
     \si_1^x \ket{\pm ; {\bf n}} = \pm \ket{\pm ;{\bf n}}
\end{equation}
for all admissible $M$, leading to 
\begin{equation} \label{eq:BlockDiagonalizationSigmaX}
     \bra{\mu_{M'}(\bfvec{q})} \si_1^x  \ket{\mu_M(\bfvec{p})}=0, \quad \mbox{if $M \neq M'$.} 
\end{equation}

\begin{figure}[tbp]
\begin{tabular}{c}
\includegraphics[width=0.98\columnwidth]{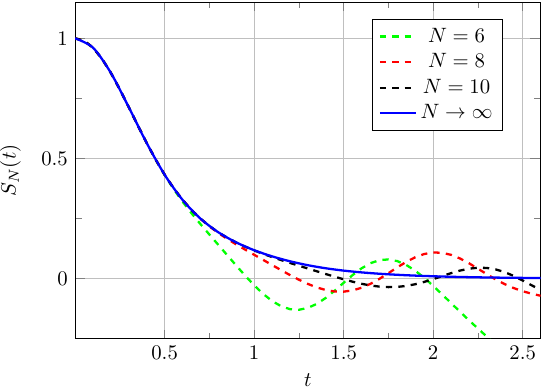}\\[12.pt]
\includegraphics[width=0.98\columnwidth]{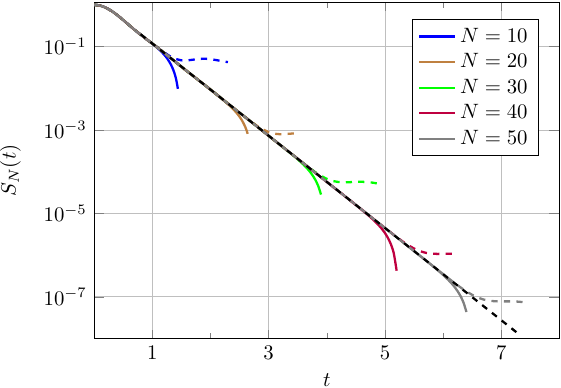}
\end{tabular}
\caption{Universal relaxation function of the spin-helix amplitude
(\ref{eq:SN(t)}) for different system sizes, in usual scale
(top panel) and in logarithmic scale (bottom panel).
\textbf{Top Panel:} Green, red and black dots correspond to
$S_6(t),S_8(t),S_{10}(t)$ respectively, while the continuous
curve shows $S(t)$ (\ref{S(t)}). The blue line is to be compared
with Fig.~2a in \cite{SHS-Ketterle}.
\textbf{Bottom Panel:} $S_N(t)$ for $N=10, 20, \ldots, 50$,
from (\ref{rep2phimn}), shows the exponential decay for large
times, given by the black dashed line, (\ref{S(t)asymptotic}).
Coloured dashed curves show $S(r,t)$ with $r=[N/4]$ from
(\ref{EqDetProduct}). Curves with the same  colour code correspond
to the same $N$.  Deviations from the straight line at large $t$
are due to finite size effects. 
}
\label{Fig-SN(t)}
\end{figure}

\begin{figure}[tbp]
\centerline{
\includegraphics[width=0.98\columnwidth]{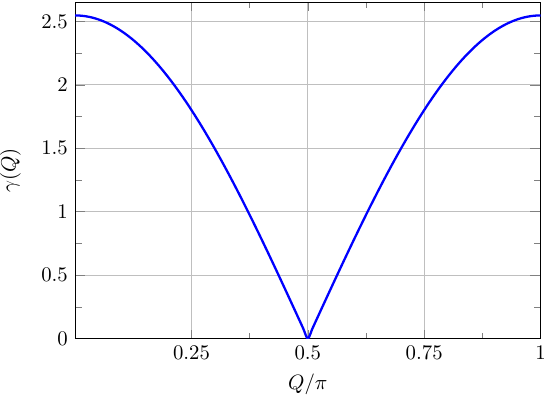}
}
\caption{Asympotic decay rate $\ga$ of the spin-helix state versus
rescaled wavevector $k=Q/\pi$, given by $\frac{8}{\pi} |\cos (\pi k)|$.
This Figure is to be compared with the experimental data shown
in Fig.~3c of \cite{SHS-Ketterle}.
}
\label{Fig-gammaDecay}
\end{figure}

Inserting  $I= \sum_{\pp,M}\ket{\mu_M(\pp)} \bra{\mu_M(\pp)}$ in
(\ref{eq:SN(t)}) and using (\ref{eq:BlockDiagonalizationSigmaX}) we obtain
\begin{multline} \label{eq:S(t)}
     S_N(t) = \sum_{\pp,\qq, M} \re^{\i (E_\pp-E_\qq) t} \\
        \times \braket{\Omega}{\mu_M(\pp)} \bra{\mu_M(\pp)} \si_1^x
	       \ket{\mu_M(\qq)} \braket{\mu_M(\qq)}{\Omega}.
\end{multline}
We also find that $\braket{\Omega} {\mu_M(\pp)}= 0$ unless $M = N/2$.
After an explicit evaluation of the matrix elements and the overlaps
and after performing the necessary summations (see \cite{Supp2}) we
eventually obtain
\begin{align}
     & S_N(t) = \real \Bigl\{\det_{m,n = 1, \dots, N/2} \Ph^{(N)}_{m,n} (t)\Bigr\},
                \label{DetPhiN}\\
     & \Ph^{(N)}_{m,n} (t) =
       \sum_{\substack{p \in B_+\\q \in B_-}}
                     \frac{(1 + \re^{- \i p})(1 + \re^{\i q})
		           \re^{\i [2(mp - nq) + t(\e_p - \e_q))]}}
		          {N^2(\re^{\i (p - q)} - 1)},\label{rep2phimn}
\end{align}
where $B_\pm$ are the sets of $p \in [- \pi, \pi)$ satisfying
$\re^{ \i p N}=\pm 1$. Eq.~(\ref{DetPhiN}) describes the relaxation
of the helix amplitude for finite periodic systems. Explicit expressions
for $S_N(t)$ for $N=4,6$ are given in \cite{Supp2}.

Analyzing the Taylor expansion of $S_N (t) = \sum_n C_n^{(N)} t^n$
at $t=0$ we observe that the Taylor coefficients $C_n^{(N)}$
stabilize for fixed $n$ and large $N$. More precisely, $C_n^{(N+2)}
= C_n^{(N)}$ for $n=0,1,\ldots, 2N-4$. Consequently, the stable
pattern gives the exact Taylor expansion about $t=0$ of the
decay of the spin-helix amplitude in the thermodynamic limit
\begin{align}
S(t) = & \lim_{N \rightarrow \infty}S_N(t),  \label{S(t)}\\
S(t) = & \,1-4 t^2 + \frac{2^5}{3}t^4 - \frac{2^6}{3}t^6 + \frac{2^9}{15}t^8 -
\frac{2^{11}}{45}t^{10} \notag \\
& + \frac{ \ 2^{12}\, 179}{14175}  t^{12}
-\frac{ 2^{16}\, 11}{14175} t^{14} + \frac{2^{16} 2987}{4465125}
t^{16} \notag \\ & - \frac{2^{18}572}{4465125} t^{18} +\ldots,
\label{eq:Taylor}
\end{align}
obtainable also by direct operatorial methods.

\textbf{Reduction to Bessel functions--}
For large $N$ the sums in (\ref{rep2phimn}) can be replaced by
integrals. Then, after some algebra, we find (see  \cite{Supp2})
that the matrix entries $\Ph^{(N)}_{m,n} (t)$ converge to
\begin{align}
& (-1)^{m-n} \Ph_{m,n} (t) = \de_{m,n} + K_{m,n} (t), \label{phirelk} \\ 
     &K_{m, n} (t) =
        \frac{t}{m - n} \bigl(J_{2m}(4t) J_{2n-1}(4t) - J_{2n}(4t) J_{2m-1}(4t)\bigr) \nonumber \\
        &+ \frac{t}{m - n} \bigl(J_{2m-1}(4t) J_{2n-2}(4t) - J_{2n-1}(4t) J_{2m-2}(4t)\bigr) \nonumber \\
        &+ \frac{\i t}{m - n - 1/2}
	  \bigl(J_{2m - 2} (4t) J_{2n} (4t) - J_{2n - 1} (4t) J_{2m - 1} (4t)\bigr)\nonumber  \\
        &- \frac{\i t}{m - n + 1/2}
	  \bigl(J_{2m - 1} (4t) J_{2n - 1} (4t) - J_{2n - 2} (4t) J_{2m} (4t)\bigr), \no\\
&K_{n,n}(t)=- (J_0(4t))^2 + (J_{2n - 1 } (4t))^2 +2 \sum_{j=0}^{2n-2}(J_j(4t))^2,
\label{eq:Kfunc}
\end{align}
where the $J_k(x)$ are Bessel functions.  
After further manipulations (see  \cite{Supp2}) and taking into account the symmetries of 
$K_{n,m}$ we finally obtain
\begin{align} \label{finaldiscretebessel}
     & S(t) = \lim_{r \rightarrow \infty} S(r,t), \\
     & S(r,t) =  \Bigl|\det_{m,n = 1, \dots, r}  A_{m,n}(t)\Bigr|^2, \,\, \label{EqDetProduct}\\ 
     & A_{m,n}(t) = \de_{m,n} +K_{m,n} (t)+ K_{m,1-n}(t).
\end{align}
These formulae represent $S(t)$ as a product of two infinite
determinants. Infinite determinants \cite{Simon77} may define
functions in very much the same way as series or integrals. As
in the present case, they may be extremely efficient in computations 
\cite{Bornemann10}. With a few lines of Mathematica code we
obtain, e.g., $S(t = 50) = 7.64483 \times 10^{- 56}$ within a
few seconds on a laptop computer. Unlike the Taylor
series (\ref{eq:Taylor}), the determinant representation determines
$S(t)$ for \textit{all} times. The function $S(t)$ shown in
Fig.~\ref{Fig-SN(t)} is directly  comparable  with the experimental
data, Fig.~2a in \cite{SHS-Ketterle}. 

Even though the true thermodynamic limit is given by
$r\rightarrow \infty$ in (\ref{finaldiscretebessel}), already
for $r=1$, when the matrix $A$ is a scalar, the function 
\begin{equation}
     S(1,t) = g_0^2 + 4 t^2\left(g_0 + \frac{g_1}{3} \right)^2,\
              g_n = J_n^2(4t) + J_{n+1}^2(4t)
\end{equation}
approximates $S(t)$ for $0\leq t \leq 0.5$ (data not shown), and
also reproduces the asymptotic Taylor expansion (\ref{eq:Taylor})
up to the order $t^7$.

Choosing $r=4$ in (\ref{EqDetProduct}) reproduces $S(t)$ with
accuracy $|S(4,t)-S(t)|<10^{-5}$, for $t<2$ which is enough for
any practical purpose. Indeed at $t=t_{max}=2$, the amplitude
$S(t)$ drops by two orders of magnitude with respect to the
initial value, $S(t_{max})\approx 0.0093 <S(0)/100$. For larger
$t$, $S(t)$ is well approximated by the asymptotics
(\ref{S(t)asymptotic}).

In addition, our numerics suggests a simple asymptotics for
$\det A(t)$,  namely 
\begin{align} 
     & \det A(t) \rightarrow  a_0   \re^{2 \i t} \re^{-\frac{4}{\pi}t },
       \quad t\gg 1,  \label{Asymp} \\
     & a_0 = 1.2295 \pm 2\times 10^{-5}. \no
\end{align}
The data were obtained by analyzing $ \det A(t)$ for $r\leq 170$, 
and for times $t< t_{m}(r) = r/2.2 -0.19$, data shown in \cite{Supp2}.  
Eq.(\ref{Asymp})  corresponds to the $S(t)$ asymptotics  
\begin{equation} \label{S(t)asymptotic}
     \lim_{t \rightarrow \infty} S(t) \approx 1.5117 \re^{-\frac{8}{\pi}t }.
\end{equation}

Using that $S(t)$ is even \cite{2023SHSdecay} we readily get
the spin-helix state decay rate from the asymptotics (\ref{S(t)asymptotic})
and the self-similarity (\ref{eq:Sx}):
\begin{equation}
     \ga(Q)= -\lim\limits_{t\rightarrow \infty}
        (t^{-1}{\log \langle \si_{n}^x(t)\rangle_Q}) = \frac{8}{\pi} |\cos (Q )|,
\end{equation}
shown in Fig.~\ref{Fig-gammaDecay} and directly comparable with the
experimental result, Fig.~3c of \cite{SHS-Ketterle}.

\textbf{Conclusions--}
In this work we propose a chiral qubit basis that possesses
topological properties while retaining a simple factorized
structure and orthonormality. The chiral basis at every site
is represented by a  pair of mutually orthogonal qubit states
and  can be implemented with usual  binary code registers.
We demonstrate the effectiveness of the chiral basis by applying
it to an experimentally relevant physical problem. Our results
in Figs.~\ref{Fig-SN(t)} and \ref{Fig-gammaDecay} are comparable
to the experimental data.

We discovered a universal function $S(t)$ that governs the
relaxation of transversal spin helices with arbitrary
wavelengths in an infinite system under XX dynamics. We obtained
the explicit determinantal form (\ref{finaldiscretebessel}) of $S(t)$
and calculated its Taylor expansion (\ref{eq:Taylor}) and its
large-$t$ asymptotics (\ref{S(t)asymptotic}). The possibility
to express correlation functions in determinantal form is typical
of integrable systems, see e.g.\
\cite{Lenard64,1993ColomoXX0basis,KoSl90,GKS20a,2021Goehmann,2021Suzuki}. 
We also obtained explicit expressions for the spin-helix state
relaxation of finite systems of qubits (\ref{DetPhiN}) that may
be useful to interpret future experiments with ring-shaped atom
arrays \cite{RingShapedArrays}, where periodic boundary conditions
can be realized.

The chiral basis can be used to diagonalize any other
Hamiltonian that commutes with the winding number operator $V$,
Eq.~(\ref{eq:Uoper}). An important example is the anisotropic
XY Hamiltonian for which we expect to be able
to obtain efficient formulae for the overlaps and, most likely,
also for the relaxation of spin helices. As for the construction
of the chiral basis, a generalization to the XYZ case has
been put forward in parallel to this work by three of the
authors and was recently published in \cite{ZKP24}.

\begin{acknowledgments}
Financial support from Deutsche Forschungsgemeinschaft through DFG project KL
645/20-2 is gratefully acknowledged. V.~P.~acknowledges support by the
European Research Council (ERC) through the advanced Grant
No.\ 694544—OMNES. X. Z. acknowledges financial support from the National
Natural Science Foundation of China (No.\ 12204519).
\end{acknowledgments}

\bibliographystyle{apsrev4-1}
\bibliography{ChiralBasis}

\clearpage

\setcounter{table}{0}
\renewcommand{\thetable}{S\arabic{table}}%
\setcounter{figure}{0}
\renewcommand{\thefigure}{SM\arabic{figure}}%
\setcounter{equation}{0}
\renewcommand{\theequation}{S\arabic{equation}}%
\setcounter{page}{1}
\renewcommand{\thepage}{SM-\arabic{page}}%
\setcounter{secnumdepth}{3}
\setcounter{section}{0}
\renewcommand{\thesection}{\arabic{section}}%
\setcounter{subsection}{0}
\renewcommand{\thesubsection}{\arabic{section}.\arabic{subsection}}%
\renewcommand{\thesection}{S-\Roman{section}}

\onecolumngrid

\section*{Supplemental Material}
This Supplemental Material contains four sections. In \ref{S-I} we prove the Theorem.
In \ref{app:SHSunderXX} we give general properties of spin-helix state (SHS)
observables including the scaling relation. In \ref{S-II} we find the
relaxation function for a finite periodic chain. In \ref{app:Bessel} we find
the relaxation function for an infinite system using Bessel functions.

\section{Proof of the Theorem}\label{S-I}
The fact that XX Hamiltonian can be diagonalized within blocks with a
fixed number of kinks $M$ follows from the commutativity of $H$ with $V$.

\noindent $\bm{M=1}$ \textbf{case.}
When $N=4m+2,\, m\in\mathbb{N}$ (only in this case $M=1$ is admissible),
we obtain the following relations
\begin{align}
\begin{aligned}
&H\ket{u;n}=2\ket{u;n-1}+2\ket{u;n+1},\quad 2 \leq n\leq N-1,\\
&H\ket{u;1}=-2\ket{u+2;N}+2\ket{u;2},\\
&H\ket{u;N}=-2\ket{u-2;1}+2\ket{u;N-1}.
\end{aligned}
\end{align}
The $2N$ eigenstates of the Hamiltonian can be expanded as follows  
\begin{align}
&&\ket{\mu_1(p)}=\sum_{m=0,2}\,\sum_{n=1}^Nf_{u+m;n}\ket{u+m;n}\quad 
\mbox{with}\quad H\ket{\mu_1(p)}=4\cos p\ket{\mu_1(p)}.
\end{align}
Then the eigen equation gives the following relations
\begin{align}
\begin{aligned}\label{FR;r1}
&2\cos p\,f_{u,n}=f_{u,n+1}+f_{u,n-1},\,\, n=2,\dots,N-1,\\
&2\cos p\,f_{u,1}=f_{u,2}-f_{u+2,N},\\
&2\cos p\,f_{u,N}=f_{u,N-1}-f_{u-2,1}.
\end{aligned}
\end{align}
Introduce the following ansatz 
\begin{align}
f_{u,n}=\re^{\i np-\frac{\i\pi\alpha u}{2}}.
\end{align}
To satisfy the above functional relations (\ref{FR;r1}), we need 
\begin{align}
-f_{u+2,N+n}=f_{u,n},\quad f_{u+4,n}=f_{u,n}.
\end{align}
which gives the twisted Bethe ansatz equations
\begin{align}
&\re^{\i Np}=-\re^{\i\al\pi},\qquad 
\re^{2\i\alpha\pi}=1.\label{BAE;r1}
\end{align}
As we have $\alpha=0,1$, the exact solutions of the BAE (\ref{BAE;r1}) are thus 
\begin{align}
&\alpha=1,\quad p=\frac{2m\pi}{N},\quad m=1,\ldots,N,\\
&\alpha=0,\quad p=\frac{(2m-1)\pi}{N},\quad m=1,\ldots,N.
\end{align}

\noindent $\bm{M=2}$ \textbf{case.}
The block with $M=2$ exists only for $N=4m$.   
Applying $H$ to two-kink states we obtain:
\begin{align}
\begin{aligned}
&H\ket{u;n_1,n_2}=2(1-\delta_{n_1+1,n_2})\ket{u;n_1+1,n_2}+2(1-\delta_{n_1+1,n_2})\ket{u;n_1,n_2-1}\no\\
&\hspace{2.2cm}+2\ket{u;n_1-1,n_2}+2\ket{u;n_1,n_2+1},\quad 2\leq n_1<n_2\leq N-1,\\
&H\ket{u;1,n_2}=2(1-\delta_{2,n_2})\ket{u;2,n_2}+2(1-\delta_{2,n_2})\ket{u;1,n_2-1}\no\\
&\hspace{2.2cm}+2\ket{u+2;n_2,N}+2\ket{u;1,n_2+1},\qquad 2\leq n_2\leq N-1,\\
&H\ket{u;n_1,N}=2\ket{u-2;1,n_1}+2\ket{u;n_1-1,N}\no\\
&\hspace{2.2cm}+2(1-\delta_{n_1+1,N})\ket{u;n_1,N-1}+2(1-\delta_{n_1+1,N})\ket{u;n_1+1,N},\quad 2\leq n_1\leq N-1,\\
&H\ket{u;1,N}=2\ket{u;2,N}+2\ket{u;1,N-1}.
\end{aligned}
\end{align}
The $2\binom{N}{2}$ eigenstates of $H$ belonging to the chiral block with $M=2$ are expanded as
\begin{align*}
&\ket{\mu_2(p_1,p_2)}=\sum_{m=0,2}\,\sum_{n_1<n_2}f_{u+m,n_1,n_2}\ket{u+m;n_1,n_2},\\
&H\ket{\mu_2(p_1,p_2)}=4\sum_{j=1}^2\cos(p_j)\ket{\mu_2(p_1,p_2)}.
\end{align*}
The eigen equation of $H$ gives the following functional relations 
\begin{align}
&[2\cos(p_1)+2\cos(p_2)]f_{u,n_1,n_2}\no\\
&=f_{u,n_1-1,n_2}+f_{u,n_1,n_2+1}+(1-\delta_{n_1+1,n_2})f_{u,n_1+1,n_2}+(1-\delta_{n_1+1,n_2})f_{u,n_1,n_2-1},\quad 1< n_1<n_2< N, \label{FR;r2;1}\\
&[2\cos(p_1)+2\cos(p_2)]f_{u,1,n_2}\no\\
&=f_{u,1,n_2+1}+f_{u+2,n_2,N}+(1-\delta_{2,n_2})f_{u,2,n_2}+(1-\delta_{2,n_2})f_{u,1,n_2-1},\quad 1< n_2<N,\label{FR;r2;2}\\
&[2\cos(p_1)+2\cos(p_2)]f_{u,n_1,N}\no\\
&=f_{u,n_1,N-1}+f_{u-2,1,n_1}+(1-\delta_{n_1+1,N})f_{u,n_1+1,N}+(1-\delta_{n_1+1,N})f_{u,n_1,N-1},\quad 1<n_1<N,\label{FR;r2;3}\\
&[2\cos(p_1)+2\cos(p_2)]f_{u,1,N}=f_{u,1,N-1}+f_{u,2,N}.\label{FR;r2;4}
\end{align}
Employ the ansatz 
\begin{align}
f_{u,n_1,n_2}=\re^{-\frac{\i\al u\pi}{2}}\sum_{a,b}A_{a,b}\,\re^{\i (n_ap_a+ n_bp_b)}.
\end{align}
where $\{a,b\}=\{1,2\}$.
To satisfy Eq.~(\ref{FR;r2;1}), we get the ``two-body scattering matrix"
\begin{align}
\frac{A_{2,1}}{A_{1,2}}\equiv -1.\label{S;r2}
\end{align}
The compatibility of Eqs.~(\ref{FR;r2;2}), (\ref{FR;r2;3}) requires 
\begin{align}
f_{u,m,n}=f_{u+2,n,m+N},\quad f_{u,m,n}=f_{u+4,m,n},
\end{align}
which leads to the Bethe ansatz equations
\begin{align}
\begin{aligned}\label{BAE;r2}
&\re^{\i Np_j}= -\re^{\i\alpha\pi},\quad j=1,2,\\
&\re^{2\i\alpha\pi}=1.
\end{aligned}
\end{align}
Solutions of BAE (\ref{BAE;r2}) are 
\begin{align}
&\alpha=1,\quad p_1=\frac{2j_1\pi}{N},\quad p_2=\frac{2j_2\pi}{N}\quad 1\leq j_1<j_2\leq N,\\
&\alpha=0,\quad p_1=\frac{(2j_1-1)\pi}{N},\quad p_2=\frac{(2j_2-1)\pi}{N} \quad 1\leq j_1<j_2\leq N.
\end{align}

\noindent \textbf{Arbitrary $\bm M$ case.}
The eigenstates $\ket{\mu_M(\pp)}$ can be expanded as
\begin{align}
\ket{\mu_M(\pp)}=\sum_{m=0,2}\,\sum_{\substack{n_1<n_2<\ldots\\\cdots<n_M}}f_{u+m,n_1,\ldots,n_M}\ket{u+m;n_1,\ldots,n_M}.\label{BA;M}
\end{align}
Employ the ansatz 
\begin{align}
f_{u,n_1,\ldots,n_M}=\re^{-\frac{\i\al \pi
  u}{2}}\sum_{r_1,\ldots,r_M}A_{r_1,\ldots,r_M}\re^{\i\sum_{k=1}^M n_{r_{k_1}}p_{r_{k_1}}},\label{Amplitude;M}
\end{align}
where $\{r_1,\ldots,r_M\}=\{1,2,\ldots,M\}$.
The amplitudes $A_{\ldots}$ satisfy 
\begin{align}
\frac{A_{\ldots,r_k,r_{k+1},\ldots}}{A_{\ldots,r_{k+1},r_k,\ldots}}= -1.\label{S;rM}
\end{align}
The Bethe ansatz equations read
\begin{align}
\begin{aligned}\label{BAE;rM}
	&\re^{\i Np_j}=-\re^{\i\alpha\pi},\quad j=1,\ldots,M,\\
	&\re^{2\alpha\pi}=1.
\end{aligned}
\end{align}
The solutions of BAE (\ref{BAE;rM}) are
\begin{align}
	&\alpha=1,\quad p_k=\frac{2j_k\pi}{N},\quad k=1,\ldots,M,\quad 1\leq j_1<j_2<\ldots<j_M\leq N,\\
	&\alpha=0,\quad p_k=\frac{(2j_k-1)\pi}{N},\quad k=1,\ldots,M, \quad 1\leq j_1<j_2<\ldots<j_M\leq N.
\end{align}
The energy in terms of the Bethe roots $\pp=\{p_1,\ldots,p_M\}$ is 
\begin{align}
E_{\pp}=4\sum_{j=1}^M\cos(p_j).
\end{align}

\textit{Remark}: The eigenstates constructed in (\ref{BA;M}), (\ref{Amplitude;M})
and (\ref{S;rM}) are consistent with the ones in the main text. By normalizing
$\ket{\mu_M(\pp)}$ in (\ref{BA;M}) we arrive at Eq.~(13).

\bigskip

\noindent
\textbf{Proof of normalization of $\bm{\ket{\mu_M(\pp)}}$ in (13).}
For real $u$ the eigenvectors $\ket{\mu_M(\pp)}$ are orthogonal, $\braket  {\mu_M(\pp)}  {\mu_{M'}(\pp')} =\de_{\bfvec{p},\bfvec{p'}}  \de_{M,M'}$.
Indeed, for $\bfvec{p}=\bfvec{p'}, M=M'$ we have
\begin{align}
&\braket  {\mu_M(\pp)}  {\mu_{M}(\pp)}=\no\\
&= \frac{1}{2N^{M}(M!)^2}\ \sum_{\mathbf{n},\bfvec{m}} T_{\bfvec{n}}\, T_{\bfvec{m}}  \sum_{Q,Q'} (-1)^{Q+Q'} \re^{-\i \sum_{j=1}^M n_j  p_{Q_j} } \re^{\i \sum_{j=1}^M m_j  p_{Q'_j} } \nonumber\\
&\quad \times \left(
\bra{u;n_1,n_2,\ldots,n_M}-\re^{-\i p_1 N}   \bra{u+2;n_1,n_2,\ldots,n_M}\right)
\left(\ket{u;m_1,m_2,\ldots,m_M}-\re^{\i p_1 N}   \ket{u+2;m_1,m_2,\ldots,m_M}\right)\no\\
&=\frac{1}{N^MM!}\sum_{n_1,n_2,\ldots} T_{\bfvec{n}}\, T_{\bfvec{n}} \,\sum_{Q,Q'} (-1)^{Q+Q'}
\re^{-\i \sum_{j=1}^M n_j  p_{Q_j} } \re^{\i \sum_{j=1}^M n_j  p_{Q'_j}}.
\end{align}
In the last passage, we used the orthogonality of the vectors $\ket{u;m_1,m_2,\ldots, m_M}$, $\ket{u+2;m_1,m_2,\ldots ,m_M}$, and assumed \textit{real}
$u$, so that $\braket{u;m_1,m_2,\ldots,m_M} {u;m_1,m_2,\ldots, m_M}=1$.  In
the last sum, only $Q'=Q$ permutations survive, there are $M!$ such permutations,
so we continue:
\begin{align}
\braket  {\mu_M(\pp)}  {\mu_{M}(\pp)}=
\frac{1}{N^M}\sum_{n_,n_2,\ldots,n_M} 1  = 1.
\end{align}

\section{General properties of SHS observables under XX evolution}
\label{app:SHSunderXX}
\noindent
\textbf{Relation  between spatially shifted observables.}
Let $A_{n}$  and  $A_{n+1}$ be the same operators  shifted by one site.   Let
\begin{align}
&\langle A_n(t) \rangle_Q   = \bra{\Psi_Q} \re^{\i H t}  A_n  \re^{-\i H t} \ket{\Psi_Q},\\
&\ket{\Psi_{Q}}=\frac{1}{\sqrt{2^N}}\bigotimes_{n=0}^{N-1} 
\binom {\re^{-\i \frac{Qn}{2} }} {\re^{\i\frac{Qn}{2}} } \label{eq:SHS}, \\
& QN =0 \mod \ 2\pi.
\end{align}
 Then, 
\begin{align}
&\langle A_{n+1}\rangle= \langle W_Q^\dagger  A_{n} W_Q\rangle, \label{eq:shift}\\
&W_Q = \bigotimes_{n}   \re^{-\i  \frac{Q}{2} \si_n^z}, \nonumber
\end{align}
where we omit the explicit dependence on $t$, since the equation is valid for any $t$. 

For a proof let $T$ be the operator inducing a shift by one lattice site to the right.
Obviously, $[T,H]= [W_Q,H] =0$. Using the easily verifiable relation
\begin{align*}
&T \ket{\Psi_Q} =  W_Q \ket{\Psi_Q},
\end{align*}
we obtain 
\begin{align*}
& \langle A_{n+1}(t)\rangle = \bra{\Psi_Q} T^\dagger \re^{\i H t}    A_n \re^{-\i H t} T \ket{\Psi_Q} \\
& = \bra{\Psi_Q}   W_Q^\dagger  \re^{\i H t} A_n\re^{-\i H t}  W_Q \ket{\Psi_Q}\\
&= \bra{\Psi_Q}    \re^{\i H t}  W_Q^\dagger   A_n  W_Q  \re^{-\i H t} \ket{\Psi_Q}, 
\end{align*}
i.e.  (\ref{eq:shift}).  Specifying $A_n = \sigma^+_n$ and   $A_n = \sigma^{-}_n$ and applying  (\ref{eq:shift})
we obtain  

\begin{align}
& \langle \si_{n+1}^\pm(t) \rangle_Q= \re^{\pm \i Q } \langle \si_n^\pm(t) \rangle_Q. \label{eq:ShiftTheorem}
\end{align}
This implies 
\begin{align}
& \langle \si_{n}^x(t)\rangle_Q= S_N(Q,t) \cos (Qn-\alpha(t)),    \label{eq:SxApp}\\
& \langle \si_{n}^y(t)\rangle_Q= S_N(Q,t) \sin (Qn-\alpha(t)),   \label{eq:SyApp} \\
& \langle \si_{n}^z(t)\rangle_Q=0,\qquad  \forall n, \label{eq:Sz}
\end{align}
where $S_N(Q,t)$ is a helix amplitude and $\alpha(t)$ is a possible Galilean shift.

\bigskip

Further, consider operators $U_x= \bigotimes_{n=1}^N \si_n^x$ and  
\begin{align}
U_Q= \re^{-\i\frac{Q}{2} \sum_{n=1}^{N} n \, \si_n^z} \label{app:UQ}
\end{align}
which have the properties
\begin{align}
&U_x \ket{\Psi_{Q}}= \ket{\Psi_{-Q}},\\
&U_Q^2 \ket{\Psi_{Q}}= \ket{\Psi_{-Q}}.
\end{align}
Using the  $[H,U_x]=[H,U_Q]=0$, $U_x^\dagger = U_x$, we obtain
\begin{align}
&\langle{A}\rangle_{Q }= \langle U_x A U_x\rangle_{-Q},\\
&\langle{A}\rangle_{Q }= \langle U_Q^{-2} A U_Q^2\rangle_{-Q}\label{App:Qprop}
\end{align}
In particular for one-point correlations one obtains

 \begin{align}
&\langle\si_n^x\rangle_{Q}= \langle \si_n^{x} \rangle_{-Q}, \label{eq:UxActionX}\\
&\langle{\si_n^{y,z}}\rangle_{Q}= -\langle {\si_n^{y,z}}\rangle_{-Q}. \label{eq:UxActionXY}
\end{align}
On the other hand,  from (\ref{App:Qprop}) we obtain for $A \equiv \si_N^\alpha$:

\begin{align}
&\langle{\si_N^{y}}\rangle_{Q}=\langle{\si_N^{y}}\rangle_{-Q} ,\label{eq:UxActionXY1}
\end{align}
leading to $\langle{\si_N^{y}}\rangle_{Q}=0$ and  $\alpha(t) =0$ in (\ref{eq:SxApp}),(\ref{eq:SyApp}).  

\bigskip
Further, we prove the scaling property
\begin{align}
&S_N(Q,t) = S_N(0,t \cos Q) \label{App:self-similarity}
\end{align}
Take a unitary operator $U_Q$ from (\ref{app:UQ}) satisfying
$\ket{\Psi_{Q}} = U_Q \ket{\Psi_{0}}$,  and let $A' = U_Q^\dagger A \, U_Q$ for  an arbitrary operator $A$. Then
\begin{align}
& \langle A(t)\rangle_Q=\bra{\Psi_{0} }\re^{\i H't} A' \re^{-\i H't} \ket{\Psi_{0}}. \nonumber
\end{align}
To calculate $H'=2 \sum_n (\si_n^+ \si_{n+1}^- + h.c.)'$ we substitute
$(\si_n^\pm)'=\re^{\pm i n Q} \si_n^\pm$,  and obtain
\begin{align}
&H' = H \cos Q +
\frac{\sin Q}{2} J,\label{eq:H'}\\
&J=4 i \sum_n ({\sigma}_n^+\sigma_{n+1}^- - h.c.). \nonumber
\end{align}
For the XX  Hamiltonian 
the total magnetization current $J$  is a constant of motion:
$[H,J]=0$.  Using, in addition, an easily checkable $J \ket{\Psi_{0}} =0$,  we  obtain
$\re^{-\i H't} \ket{\Psi_{0}}= \re^{-\i Ht \cos Q }\ket{\Psi_{0}}$,
i.e.
\begin{align}
& \langle A(t)\rangle_Q= \langle A'(t \cos Q)\rangle_0\label{eq:HxxTheorem}
\end{align}
Eq.~(\ref{eq:HxxTheorem}) leads to is self-similarity 
(\ref{App:self-similarity}),  and consequently to Eq.~(16).

\section{Derivation of determinantal formula (21)}
\label{S-II}
 
Here we calculate the relaxation of the SHS magnetization profile given by
\begin{align}
&S_N(t) = \sum_{\pp,\qq,  \mbox{ \,all}  \, M} \braket{\Omega}{\mu_M(\pp)} \re^{\i (E_\pp-E_\qq) t} \bra{\mu_M(\pp)} \si_1^x \ket{\mu_M(\qq)} \braket{\mu_M(\qq)}{\Omega}, \label{app:SN(t)}
\end{align}
using the chiral basis (13) with
\begin{align}
& u\equiv u_0 = 1. \label{u0}
\end{align}
First,  one can verify straightforwardly that all nonzero overlaps $\braket{\Omega} {\mu_M(\pp)}$  
belong to a single block with 
\begin{align}
& M \equiv \frac{N}{2} \label{M0}
\end{align}
kinks.  Consequently,   (\ref{app:SN(t)}) simplifies:
\begin{align}
&S_N(t) = \sum_{\pp,\qq} \braket{\Omega}{\mu_M(\pp)} \re^{\i (E_\pp-E_\qq) t} \bra{\mu_M(\pp)} \si_1^x \ket{\mu_M(\qq)} \braket{\mu_M(\qq)}{\Omega}, \label{app:SN2}
\end{align}
where here and in the following we assume (\ref{M0}).
In the following we demonstrate that  both nonzero overlaps $\braket{\Omega} {\mu_M(\pp)}$ and $\bra{\mu_M(\pp)} \si_1^x \ket{\mu_M(\qq)} $ in (\ref{app:SN(t)})
can be expressed as  $M \times M$ determinants.

\bigskip \noindent
\textbf{Overlaps.} The overlaps are
\begin{align}
\braket{\Omega}{\mu_M(\pp)}&=\frac{1}{\sqrt{2 N^{M}}} \bra{\Omega} \sum_{Q}\sum_{n_j<n_{j+1}} (-1)^Q 
\re^{\i \sum_{k} p_{Q_k} n_k} 
\left(
\ket{u_0;n_1,n_2,\ldots n_{M}} - \re^{\i p_1 N} \ket{u_0+2;n_1,n_2,\ldots n_{M}}
 \right)\no\\
&=\frac{1}{\sqrt{2 N^{M}}}\sum_{Q} \sum_{n_j<n_{j+1}}(-1)^Q 
\re^{\i \sum_{k} p_{Q_k} n_k} 
\braket {\Omega} {u_0;n_1,n_2,\ldots n_{M}}, \label{eq:overlapSHS-MuP}
\end{align}
where in the passage from the first to the second line we used
$\braket{\Omega}{u_0+2;n_1,n_2,\ldots n_{M}}=0$ for all values of $n_j$.
Next, we  establish that for the allowed values of $n_j$, i.e.  $1\leq
n_1<n_2< \ldots <n_{M}\leq N$ we have
\begin{align}
&\braket {\Omega} {u_0;n_1,n_2,\ldots n_{M}}= \frac{K_N}{\sqrt{2^N}}
(\de_{n_1,1} + \de_{n_1,2}) (\de_{n_2,3} + \de_{n_2,4}) \cdots(\de_{n_{M},N-1} + \de_{n_{M},N}),
\label{eq:overlapSHS-basis}
\end{align}
where $K_N$ depends on the system size only,
\begin{align}
K_N=(1-\i)^{\frac{N}{2}} (-\i)^{\frac{N^2}{4}}. 
\end{align}

Substituting (\ref{eq:overlapSHS-basis}) into 
(\ref{eq:overlapSHS-MuP}) we obtain
\begin{align}
&\braket{\Omega}{\mu_{M}(\pp)} =\frac{K_N}{\sqrt{2^{N+1} N^{M}}}\sum_{Q} (-1)^Q \sum_{n_j<n_{j+1}}
\re^{\i \sum_{k=1}^{M} p_{Q_k} n_k} \prod_{j=1}^{M} (\de_{n_j,2j} + \de_{n_j,2j-1})\no\\
&\hspace{1.75cm}=\frac{K_N}{\sqrt{2^{N+1} N^{M}}}\sum_{Q}  (-1)^Q g_1(p_{Q_1})  g_2(p_{Q_2}) \ldots  g_{M}(p_{Q_{M}})\no\\
&\hspace{1.75cm}=\frac{K_N}{\sqrt{2^{N+1} N^{M}}}\det_M[G(\pp)],    \label{app:detG} \\
&G_{k,m}(\pp) = g_k(p_m)  = \sum_{n_k=2k-1}^{2k} \re^{\i  p_m n_k}= 
\re^{2\i k p_m} \left( 1+ \re^{-\i p_m}  \right).
\label{res:overlapSHS-MuP}
\end{align}

\noindent
\textbf{Calculation of $\bm{\bra{\mu(\pp)} \si_1^x \ket{\mu(\qq)}}$.}
First, note that with the choice (\ref{u0}) all basis states (13) become eigenstates
of $\si_1^x$, with eigenvalues $+1$ or $-1$,
\begin{align}
& \si_1^x \ket{u_0; n_1,n_2, \ldots} =  \ket{u_0; n_1,n_2, \ldots},  \label{eq:sxEigen}\\   
& \si_1^x \ket{u_0+2; n_1,n_2, \ldots} = -\ket{u_0+2; n_1,n_2, \ldots}.
\end{align}

For $M=1$  the XX eigenfunctions in the block with $M=1$ are $\ket{\mu_1(p)} = \frac{1}{\sqrt{2N}} \sum_{n=1}^N \re^{\i p n } (\ket{u_0;n}- \re^{\i p N} \ket{u_0+1;n})$ and  we get
\begin{align}
&s_{p',p}=\bra{\mu_1(p')} \si_1^x \ket{\mu_1(p)} = \frac{1}{2N} \sum_{n=1}^N \re^{\i (p-p')n} (1 - \re^{\i(p-p')N} ).
\end{align}
The factor $(1 - \re^{\i (p-p')N})=0$ if $p$ and $p'$ have the same parity. If the parities 
of $p,p'$ are different, i.e. $\re^{\i (p-p')N} =-1$, we obtain
\begin{align}
&s_{p',p}=\frac{1}{N} \sum_{n=1}^N \re^{\i (p-p')n} =\frac{2}{N}\frac{1}{\re^{\i (p'-p)}-1}.  \label{eq:spp}
\end{align}

For arbitrary $M$ after some algebra we obtain 
\begin{align}
&\bra{\mu_M(\pp)} \si_1^x \ket{\mu_M(\qq)} =0, \quad  \mbox{if} \ \     \re^{\i(p_1-q_1)N}=1,\\
&\bra{\mu_M(\pp)}  \si_1^x \ket{\mu_M(\qq)} =\sum_{Q} (-1)^Q \prod_{j=1}^M  s_{p_j, q_{Q_j}}=
M^{-M}\det_M [F(\bfvec{p},\bfvec{q})], \quad \mbox{if} \ \ \re^{\i (p_1 - q_1)N}=-1, \label{app:detF} \\
&F_{n,m}(\bfvec{p},\bfvec{q})= \frac{1}{\re^{\i (p_n-q_m)}-1}.
\end{align}

Substituting (\ref{app:detF}) and   (\ref{app:detG})  into (\ref{app:SN2}) we finally obtain
\begin{align}
&S_N(t) = \frac{1}{N^N}   \sum_{\pp  \, \mbox{even}} \sum_{\qq  \, \mbox{odd}} \cos ((E_\pp-E_\qq) t) \ 
\det G(\pp) \det G(-\qq) \det F(\pp,\qq).
\label{eq:S(t)-2}
\end{align}
Here we used the fact that contributions of  $\pp,\qq$ being of even/odd parity,
and being of odd/even parity in (\ref{app:SN2}) are  identical.

Product forms for the determinants  $\det F,\, \det G$ are obtained
using the well-known expressions for Vandermonde and Cauchy determinants,
\begin{align}
&\det G(\pp) =
\re^{\i \,\frac{N+1}{2}  \sum_{k} p_k} \ 2^{\frac{M(M+1)}{2}} \ (-\i)^{\frac{M(M-1)}{2}}
\prod_{1\leq n_1<n_2\leq M} \sin (p_{n_1}-p_{n_2}) \,
 \prod_{n=1}^{M}  \cos\frac{p_{n}}{2}, 
\label{res:SHSoverlap}\\
&\det F(\pp,\qq) =2^{-M} A_N \  \re^{-\frac{\i}{2} \sum_{k=1}^{M} (p_k-q_k)} \prod_{1\leq n_1<n_2\leq M} \sin\frac{p_{n_1}-p_{n_2}}{2} \,   \sin\frac{q_{n_1}-q_{n_2}}{2} 
\prod_{n,m=1}^{M} \csc\frac{p_{n}-q_{m}}{2},
\label{res:SigmaXoverlap}\\
&A_N =
\left\{
\begin{array}{ccc}
&1 , &\quad \frac{N}{2} \ \mbox {even},\\[4pt]
&-\i, & \quad \frac{N}{2} \ \mbox {odd}.
\end{array}
\right.
 \label{res:A(N)}
\end{align}

Substituting the above into (\ref{eq:S(t)-2}), we get explicit expressions for
the decay function $S_N(t)$  for different system sizes. E.g., for $N=4, 6$ we have 
\begin{align}
&S_{4}(t) = \frac{1}{8} \left(2 \cos (4 t)+\left(3+2 \sqrt{2}\right) \cos \left(4 \left(\sqrt{2}-1\right)
   t\right)+\left(3-2 \sqrt{2}\right) \cos \left(4 \left(1+\sqrt{2}\right) t\right)\right)  \label{S4(t)}\\
&S_{6}(t) =\frac{1}{96} \left(8 \cos (4 t)+2 \cos (8 t)+4 \cos \left(4 \sqrt{3} t\right)+\left(26+15
   \sqrt{3}\right) \cos \left(4 \left(\sqrt{3}-2\right) t\right)+\right.  \nonumber   \\
& +\left.  2 \left(7+4 \sqrt{3}\right) \cos
   \left(4 \left(\sqrt{3}-1\right) t\right)+2 \left(7-4 \sqrt{3}\right) \cos \left(4
   \left(1+\sqrt{3}\right) t\right)+\left(26-15 \sqrt{3}\right) \cos \left(4 \left(2+\sqrt{3}\right)
   t\right)+2\right), \label{S6(t)}
\end{align}
while for larger $N$ the expressions for $S_N(t)$ get more bulky.

\bigskip

Further, we would like to write  the expression for $S_N(t)$ from Eq.~(\ref{eq:S(t)-2}) as 
a single determinant. To this end we first rewrite Eq.~(\ref{eq:S(t)-2}) as 
\begin{equation}
     S_N (t) = \frac{\PH_N (t) + \PH_N (-t)}{2} \epc
\end{equation}
where
\begin{align} 
\label{defphi}
    &\PH_N (t) = \frac{1}{N^N} \frac{1}{(M!)^2} \sum_{\pv \in B_+^M} \sum_{\qv \in B_-^M}
		 \re^{\i t \sum_{n=1}^M (\e(p_n) - \e(q_n))}
                 \det_M G(\pv) \det_M F(\pv, \qv) \det_M G (- \qv),\\
  &\e(p) \equiv 4 \cos p, \quad M\equiv \frac{N}{2}.
\end{align}
Here  $B_\pm$ are the sets of  $p \in [- \pi, \pi)$ satisfying $1 \mp \re^{- \i p N}=0$.

Define the row vector $\Fv (p)$ with coordinates
\begin{equation}
     \Fv (p)_n = \frac{1}{\re^{\i (p - q_n)} - 1} \epp
\end{equation}
Then 
\begin{align}
 &    \frac{1}{N^M M!} \sum_{\pv \in B_+^M} \re^{\i t \sum_{n=1}^M \e(p_n)}
        \det_M G(\pv) \det_M F(\pv, \qv)  \no \\& 
     =\frac{1}{N^M M!} \sum_{\s \in \mathfrak{S}^M} \sign(\s)
        \sum_{\pv \in B_+^M} \re^{\i t \sum_{n=1}^M \e(p_n)} 
	\re^{2 \i p_1 \s 1}(1 + \re^{- \i p_1}) \dots
	\re^{2 \i p_M \s M}(1 + \re^{- \i p_M}) 
	\det \begin{pmatrix} \Fv (p_1) \\ \vdots \\ \Fv (p_M) \end{pmatrix} \epp
\end{align}
We pull the factor $\re^{\i(2 p_j \s j + t \e(p_j)}(1 + \re^{- \i p_j})$
into the $j$th row of the determinant, rename the summation indices
$p_j \rightarrow p_{\s j}$ and use the antisymmetry of the
determinant with respect to the interchange of rows. Then
\begin{equation}
    \frac{1}{N^M M!} \sum_{\pv \in B_+^M} \re^{\i t \sum_{n=1}^M \e(p_n)}
        \det_M G(\pv) \det_M F(\pv, \qv)
    = \frac{1}{N^M} \sum_{\pv \in B_+^M}
	\det
	   \begin{pmatrix}
	   (1 + \re^{- \i p_1}) \re^{\i(2 p_1 + t \e(p_1)} \Fv (p_1) \\
	      \vdots \\
	   (1 + \re^{- \i p_M}) \re^{\i(2 M p_M + t \e(p_M)} \Fv (p_M)
	   \end{pmatrix} \epp
\end{equation}
Here the sums $\sum_{\pv \in B_+^M} =\sum_{p_1 \in B_+} \sum_{p_2 \in B_+}  \ldots
\sum_{p_M \in B_+} $
can be pulled inside the determinant, using the multilinearity of the determinant.
Introducing the column vectors $\Hv (q)$ with coordinates
\begin{equation}
     \Hv (q)^m = \frac1N \sum_{p \in B_+}
                 \frac{(1 + \re^{- \i p)} \re^{\i (2mp + t\e(p))}}
		      {\re^{\i (p - q)} - 1},
\end{equation}
we thus obtain 
\begin{equation}
     \frac{1}{N^M M!} \sum_{\pv \in B_+^M} \re^{\i t \sum_{n=1}^M \e(p_n)}
        \det_M G(\pv) \det_M F(\pv, \qv) =
        \det \bigl(\Hv (q_1), \dots, \Hv (q_M)\bigr) \epp
\end{equation}

Inserting the latter into (\ref{defphi}) results in
\begin{align}
     \PH_N (t) =&
        \frac{1}{N^M M!} \sum_{\s \in \mathfrak{S}^M} \sign(\s)
        \sum_{\qv \in B_-^M} \re^{- \i t \sum_{n=1}^M \e(q_n)} \times \no \\
	&\re^{- 2 \i q_1 \s 1}(1 + \re^{\i q_1}) \dots
	\re^{- 2 \i q_M \s M}(1 + \re^{\i q_M})
        \det \bigl(\Hv (q_1), \dots, \Hv (q_M)\bigr) \epp
\end{align}
Proceeding similarly as above, we conclude that
$\PH_N (t) = \det_M \Ph^{(N)} (t)$, where
\begin{equation} \label{rep2phimn-App}
     \Ph^{(N)}_{m,n} (t) = \frac{1}{N^2} \sum_{p \in B_+} \sum_{q \in B_-}
                     \frac{(1 + \re^{- \i p})(1 + \re^{\i q})
		           \re^{\i [2(mp - nq) + t(\e(p) - \e(q))]}}
		          {\re^{\i (p - q)} - 1},
\end{equation}
i.e., Eq.~(22).

\section{\boldmath Large $N$: from sums to Bessel functions}
\label{app:Bessel}

For large $N$ it is convenient to approximate the sums 
(\ref{rep2phimn-App}) via an integral.  Naively,
\begin{align}
     \frac{1}{N} \sum_{p \in B_+} f(p) & \approx 
	\int_{- \p}^\p \frac{\rd p}{2 \p} \: f(p). 
\end{align}
Within a more careful approach we introduce the oriented curves
\begin{equation}
     {\cal C}_\eps = [- \p, \p] - \i \eps \cup [\p, - \p] + \i \eps,
\end{equation}
where $\eps > 0$. Then we obtain for any function $f$ that is $2 \p$
periodic and holomorphic in a strip $S = \bigl\{z \in {\mathbb C}\big|
|\Im z| < 2 \eps\bigr\}$ that
\begin{align}
     \frac{1}{N} \sum_{p \in B_+} f(p) & =
        \int_{{\cal C}_\eps} \frac{\rd p}{2 \p} \: \frac{f(p)}{1 - \re^{-\i p N}}.
\end{align}
Analogically, for sums over $B_{-}$ we can choose a different $\eps' > 0$
such as to obtain 
\begin{align}
     \frac{1}{N} \sum_{q \in B_-} f(q) & =
        \int_{{\cal C}_{\eps'}} \frac{\rd q}{2 \p} \: \frac{f(q)}{1 + \re^{-\i q N}}. 
\end{align}
Denoting $f(p,q) \equiv 
(1 + \re^{- \i p})(1 + \re^{\i q}) \re^{\i [2(mp - nq) + t(\e(p) - \e(q))]}$
we can write down the sum in (\ref{rep2phimn-App}) as 

\begin{align} \label{PhNint}
     \Ph^{(N)}_{m,n} (t) &= \de_{m, n} +
	\int_{{\cal C}_\eps} \frac{\rd q}{2\p} \:
	\int_{{\cal C}_{\eps'}} \frac{\rd p}{2\p} \:
        \frac{f(p,q)}
	     {(1 - \re^{- \i pN}) (1 + \re^{- \i qN})
	      (\re^{\i (p - q)} -1) } \epc
\end{align}
where $m, n = 1, \dots, M$ and where we chose $0 < \eps' < \eps$ not necessarily
small. The Kronecker delta terms stem from the relative pole of the integrand.  
For large $N$ we get $\re^{- \i pN} \rightarrow 0$ on the lower part of ${\cal C}_\eps$
and $\re^{- \i pN} \rightarrow \infty$ on the upper part of ${\cal C}_\eps$, while
$\re^{- \i qN} \rightarrow 0$ on the lower part of ${\cal C}_{\eps'}$ and
$\re^{- \i qN} \rightarrow \infty$ on the upper part of ${\cal C}_{\eps'}$.
Moreover, due to  $\Im (p - q) = \eps - \eps' > 0$ we can represent the
denominator using the geometric series
\begin{equation}
     \frac{1}{1 - \re^{\i(p - q)}} =  \sum_{k=0}^\infty \re^{\i (p - q)k},
\end{equation}
so we get 

\begin{align} \label{tlasym}
     \Ph_{m,n} (t) 
        = \de_{m, n} +
	\int_{- \p - \i \eps}^{\p - \i \eps} \frac{\rd q}{\p} \:
	\int_{- \p - \i \eps'}^{\p - \i \eps'} \frac{\rd p}{\p} \:
f(p,q)  \sum_{k=0}^\infty \re^{\i (p - q)k}.
\end{align}
Interchanging summation and integration,  sending $\eps, \eps' \rightarrow +0 $ and recalling the definition of $f(p,q)$
we obtain
\begin{align} \label{rep3phimn}
     \Ph_{m,n} (t) &= \de_{m,n}
        - \sum_{k=0}^\infty {\cal J}_{2m + k} (t) {\cal J}_{2n + k}^* (t) \epc\\
     {\cal J}_n (t) &=
        \int_{- \p}^\p \frac{\rd p}{2 \p} (1 + \re^{- \i p}) \re^{\i [n p + t\e(p))]}
     = \i^{n} \left(J_{n}(4t) - \i J_{n-1}(4t)     \right)
\end{align}
where $J_n(x)$ is a Bessel function.  
The last passage follows from a well-known identity for 
Bessel functions:
\begin{equation} \label{besselintrep}
        \int_{- \p}^\p \frac{\rd p}{2 \p} \re^{\i [n p + t\e(p))]} = \i^n J_n (4t).
\end{equation}
Then
\begin{align} \label{firstbessel}
        (-1)^{m-n} \Ph_{m,n} (t) &= \de_{m,n} +K_{m,n}(t),\\
K_{m,n}(t) &=- \sum_{k=0}^\infty \bigl( J_{2m + k} (4t) - \i J_{2m - 1 + k} (4t) \bigr)
           \bigl( J_{2n + k} (4t) + \i J_{2n - 1 + k} (4t) \bigr) \epp
\label{app:KmnSums}
\end{align}
Note that the factor $(-1)^{m-n}$ can be dropped inside the
determinant.

Bessel functions satisfy the recurrence relations
\begin{subequations}
\label{recur}
\begin{align} \label{sumcur}
     J_{n-1} (t) + J_{n+1} (t) & = \frac{2n}{t} J_n (t) \epc \\[1ex] \label{diffcur}
     J_{n-1} (t) - J_{n+1} (t) & = 2 J_n' (t),
\end{align}
\end{subequations}
and have the following large-$n$ asymptotics for fixed $t$,
\begin{equation} \label{besselasyn}
     J_n (t) \sim \frac{(t/2)^n}{n!} \epp
\end{equation}
Using the first recurrence relation and the large-$n$ asymptotics,
it is easy to prove the identity \cite{2000-Borodin} 
\begin{equation} \label{besskern}
     \sum_{k = 0}^\infty J_{m + k} (t) J_{n + k} (t) =
          \frac{t}{2(m - n)} \bigl[J_{m-1} (t) J_n (t) - J_{n-1} (t) J_m (t)\bigr] \epc
\end{equation}
valid for all $m, n \in {\mathbb Z}$.

Using the latter relation together with (\ref{diffcur}) in
(\ref{firstbessel}) we obtain
\begin{align*}
     K_{m, n} (t) &=
        \frac{t}{m - n} \bigl(J_{2m}(4t) J_{2n-1}(4t) - J_{2n}(4t) J_{2m-1}(4t)\bigr) \\
        &+ \frac{t}{m - n} \bigl(J_{2m-1}(4t) J_{2n-2}(4t) - J_{2n-1}(4t) J_{2m-2}(4t)\bigr) \\
       & + \frac{\i t}{m - n - 1/2}
	  \bigl(J_{2m - 2} (4t) J_{2n} (4t) - J_{2n - 1} (4t) J_{2m - 1} (4t)\bigr) \\
       & - \frac{\i t}{m - n + 1/2}
	  \bigl(J_{2m - 1} (4t) J_{2n - 1} (4t) - J_{2n - 2} (4t) J_{2m} (4t)\bigr),
\end{align*}
i.e.~(26).  
The above expression for the matrix elements $ K_{m, n}$ is
valid also for the diagonal elements $K_{n,n}$ in the sense of the limit
$
K_{n,n} = \lim_{\eps \rightarrow 0} K_{n+\eps, n-\eps}
$.
This limit can be calculated explicitly in terms of Bessel functions from (\ref{app:KmnSums}):
\begin{align}
\lim_{N \rightarrow \infty}  \Ph^{(N)}_{n,n} (t)&= 1- \sum_{k = 0}^\infty \bigl( J_{2n + k} (4t) - \i J_{2n - 1 + k} (4t) \bigr)
           \bigl( J_{2n + k} (4t) + \i J_{2n - 1 + k} (4t) \bigr) \no \\
&= 1 - \sum_{k = 0}^\infty \bigl( (J_{2n + k} (4t))^2 +  (J_{2n - 1 + k} (4t))^2 \bigr) \no \\
&= 1 - 2\sum_{k = 0}^\infty (J_{2n + k} (4t))^2 -  (J_{2n - 1 } (4t))^2 \label{infSum} \\
&=1-\left(1+ J_0(4t)^2 -2\sum_{k=0}^{2n-1}(J_k(4t))^2 \right) -  (J_{2n - 1 } (4t))^2
\label{finiteSum} \\
&=- (J_0(4t))^2 + (J_{2n - 1 } (4t))^2 +2 \sum_{k=0}^{2n-2}(J_k(4t))^2,
\end{align} 
yielding the diagonal elements in Eq.~(26) of the main text.
Here the passage from (\ref{infSum}) to (\ref{finiteSum}) follows from the property
\begin{equation} \label{fromadd}
     \de_{m,n} = \sum_{k \in {\mathbb Z}} J_{k+m} (x) J_{k+n} (x) \epc
\end{equation}
which can be easily obtained using the generating function of the Bessel functions,
and from $J_{-n}(x) = (-1)^n J_n(x)$ as
\begin{align}
&2\sum_{k = 0}^\infty J_{2n + k}^2=\ \ 2\sum_{k = 2n}^\infty J_{ k}^2=\ \ 
 \sum_{k = 2n}^\infty J_{ k}^2 +   \sum_{k = -2n}^{-\infty} J_{ k}^2 
=1 - \sum_{k = -2n+1}^{2n-1} J_{ k}^2=1 + J_{ 0}^2 - 2\sum_{k = 0}^{2n-1} J_{ k}^2.
\end{align} 

Note that the elements $K_{m,n}$ satisfy 
\begin{align}
K_{m,n}^*(t)= K_{n,m}(t)=K_{m,n}(-t),   \quad \forall m,n \epp \label{app:Hermiticity}
\end{align}
It follows from (\ref{app:Hermiticity}) that any partial determinant of a matrix with elements 
(\ref{firstbessel}) is real for any real $t$,
\begin{equation}
        \real[\det_{m,n = 1 -r, \dots, M-r} \Ph_{m,n} (t)] =0,  \quad \forall t, r,
\end{equation}
leading to the obvious simplifications $\frac12 (\det [\Ph_{m,n} (t)]
+\det [\Ph_{m,n} (-t)]) =\det \Ph_{m,n} (t) $ and
\begin{align} \label{fornumerics}
     S(t) &= \lim_{N \rightarrow \infty} S_N (t)
        = \lim_{r \rightarrow \infty} S(r,t) \epc \\
S(r,t)&= (-1)^r \det_{m,n = 1 -r, \dots, r} \Ph_{m,n} (t).  \label{S(r,t)}
\end{align}

A further simplification of (\ref{fornumerics}) can be attained by using the identity 
(\ref{fromadd}). Adding and subtracting the terms with $k$ running from
$- \infty$ to $-1$ on the right hand side of (\ref{firstbessel}) and using
(\ref{fromadd}) we obtain 
\begin{align}
      &  (-1)^{m-n} \Ph_{m,n} (t) =\de_{m,n} -
        \sum_{k = 0}^\infty \bigl( J_{2m + k} (4t) - \i J_{2m - 1 + k} (4t) \bigr)
           \bigl( J_{2n + k} (4t) + \i J_{2n - 1 + k} (4t) \bigr) \nonumber \\
 &=\de_{m,n} -
        (\sum_{k = -\infty}^\infty -\sum_{k =-1}^{-\infty})\bigl( J_{2m + k} (4t) - \i J_{2m - 1 + k} (4t) \bigr)
           \bigl( J_{2n + k} (4t) + \i J_{2n - 1 + k} (4t) \bigr) \nonumber \\
&=\de_{m,n} - 2 \de_{m,n} +\sum_{k =-1}^{-\infty}\bigl( J_{2m + k} (4t) - \i J_{2m - 1 + k} (4t) \bigr)
           \bigl( J_{2n + k} (4t) + \i J_{2n - 1 + k} (4t) \bigr) \nonumber \\
&=- \de_{m,n}  +\sum_{k' =0}^{\infty} \bigl( J_{2m - k'-1} (4t) - \i J_{2m -k' -2} (4t) \bigr)
           \bigl( J_{2n - k'-1} (4t) + \i J_{2n  + k'-2} (4t) \bigr).\nonumber
\end{align}
Substituting $m=-m'+1$, $n=-n'+1$ in the above, and then using subsequently $J_{n}(z) = J_{-n}(-z)$
and $J_n(-z)=(-1)^n J_n(z)$ we obtain, in addition to $ (-1)^{m-n} \Ph_{m,n} (t) = \de_{m,n} +K_{m,n}(t)$,
\begin{equation} \label{minusphirelk}
      (-1)^{m-n} \Ph_{m,n} (t) = -\de_{m,n} - K_{1-m, 1-n} (t).
\end{equation}

 Let us denote by $X$ the antidiagonal matrix of
dimension $r$ with only entries $1$ on the antidiagonal (a
blown-up version of the Pauli matrix $\s^x$):  $X_{ij}=\de_{i+j,r+1}$.  Then (\ref{S(r,t)})
 can be written as 
\begin{align}
     &S(r,t)= (-1)^r \det_{m, n = 1-r, \dots, r} \Ph_{m, n} (t) =(-1)^r \det_{m, n = 1-r, \dots, r} (-1)^{m-n}\Ph_{m, n} (t)\\
     &=   \det \begin{pmatrix} -I_r- G_{1-r,1-r} & -G_{1-r,1} \\ G_{1,1-r} & I_r+ G_{1,1} \end{pmatrix},
\quad I_r \ \mbox{is the $r\times r$ unit matrix},
\label{compMatrix}\\
& \left(G_{a,b}\right)_{mn}= K_{a+m-1,b+n-1},\ \  n,m=1,\ldots ,r. 
\end{align}
Using (\ref{minusphirelk}),    we obtain $ -I_r- G_{1-r,1-r} = X ( I_r+ G_{1,1}) X$, while the 
antidiagonal blocks of (\ref{compMatrix}) can be written as $-G_{1-r,1} = X B$,  
$G_{1,1-r} =  B X$ where $B_{mn}= K_{m,1-n}(t)$,  with indices in the range $n,m=1,\ldots ,r$.  Denoting $G_{1,1} \equiv K$, 
with elements $K_{m,n}\equiv K_{m,n}(t)$, we rewrite  (\ref{compMatrix}) as 
\begin{align}
 S(t)&=\det \begin{pmatrix} X (I_r + K) X & X B \\ B X & I_r + K \end{pmatrix}\no\\
&= \det \begin{pmatrix} I_r + K & B \\ B & I_r + K \end{pmatrix} \no\\
        &= \det \begin{pmatrix} I_r + K + B & B+I_r + K \\
	                       B & I_r + K \end{pmatrix}\no\\
        &= \det \begin{pmatrix} I_r + K + B & 0 \\
	                       B & I_r + K - B \end{pmatrix} \no\\
&= \det(I_r + K + B) \det(I_r + K - B) \epp
\end{align}
It is possible to further simplify the last expression by noting that 
\begin{align}
&B^* = -B^t \label{Bprop} \epp
\end{align}
Indeed, from (\ref{app:Hermiticity}), (\ref{minusphirelk}) and the definition of $B$
we readily obtain that $(B^*)_{mn}= K_{m,1-n}^*=K_{n,1-m}= - K_{1-n,m} = -(B^t)_{mn}$.  
Then, using (\ref{Bprop}), we have 
\begin{align}
&\det(I + K + B)^* = \det(I + K^* + B^*) = \det(I + K^t - B^t) = \det(I + K - B) \epp
\end{align} 
It follows that 
\begin{align}
&S(t) = \left| \det(I_r + K + B) \right|^2,
\end{align}
i.e., (28).

From the above, $\det(I_r + K + B) = \re^{\i w(r,t) }
\sqrt{S(t)}$ where $w(r,t)$ is a real function. Numerically, we find  that 
\begin{align}
&w(\infty,t)=2t.
\end{align}
In particular, the function $\re^{-2\i t} \det(I_r + K + B)$ generates the correct
Taylor expansion of $\sqrt{S(t)}$ at time $t=0$, 
compatible with the Taylor expansion (24) of $S(t)$ obtained
in an alternative way. To generate more terms in the expansion, $r$ must be
increased. Empirically we find that, for given $r$, $\re^{-2\i t} \det(I_r + K + B)$
generates the correct Taylor coefficients of $\sqrt{S(t)}$ up to the order $t^{4r}$.

Finally, measuring the deviation of $w(r,t)-2t$ (the quantity which should be
vanishingly small for $t <t_m(r)$) we can estimate $t_m(r)= r/c$ and thus  the 
velocity of the light cone $c$. Another way to obtain $c$ is to track the
position of the deviation of $S(r,t)$ from the straight line (knicks in bottom
Panel of Fig.~1) in the logarithmic scale. Both approaches predict
$c\approx 2.2$ (data not shown).
 
\begin{figure}
\setlength{\tabcolsep}{16pt}
\begin{tabular}{cc}
\includegraphics[width=0.45\textwidth]{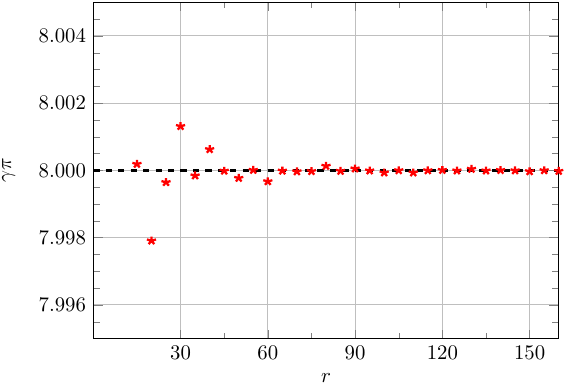} &
\includegraphics[width=0.45\textwidth]{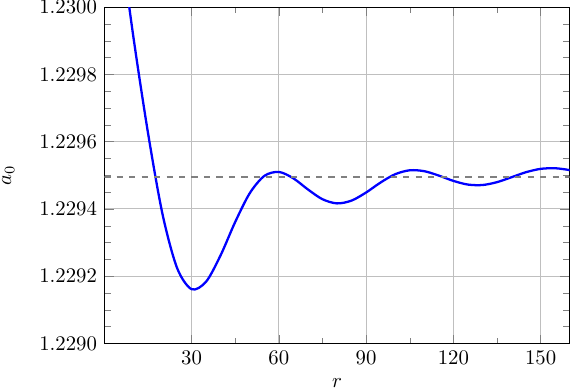}
\end{tabular}
\caption{\textbf{Left Panel:} Rescaled asymptotic decay rate $\ga \pi$ versus $r$,
where $\ga=-\lim\limits_{t\rightarrow \infty} (\partial/\partial t) {\log S(t)}$
is estimated as $\ga = (0.1t_m)^{-1} \log (S(0.9 t_m)/S(t_m) )$, where
$t_{m} =\frac{r}{2.2} - 0.19$. The dashed line indicates the predicted value $\ga \pi=8$. 
\textbf{Right Panel: } asymptotic prefactor $a_0=\sqrt{S(t_{m})} \re^{\frac{4}{\pi} t_m}$
versus $r$. The dashed line is a guide for the eye.
}
\label{Fig-Decay}
\end{figure}

Finally, from numerics we find a simple asymptotics for $\det A(t)$, namely 
\begin{align} 
 & \re^{-2\i t} \det A(t) = \sqrt{S(t)}, \quad \forall t<t_m(r) \epc \label{app:propA} \\
 & \det A(t) \rightarrow  a_0 \  \re^{2 \i t} \re^{-\frac{4}{\pi}t },  1\ll t \leq t_m(r) \epc \\
&a_0 = 1.2295 \pm 2\times 10^{-5},
\end{align}
the data obtained on the base of analyzing $ \det A(t)$ for $r<170$, see Fig.~\ref{Fig-Decay}. 
This corresponds to a $S(t)$ asymptotics of the form 
\begin{align} 
 &\lim_{t \rightarrow \infty} S(t) \approx 1.51166 \ \re^{-\frac{8}{\pi}t } \epp
\label{app:S(t)asymptotic}
\end{align}

\end{document}